# Direct determination of spin–orbit interaction coefficients and realization of the persistent spin helix symmetry

A. Sasaki[1], S. Nonaka[1], Y. Kunihashi[1], M. Kohda[1], T. Bauernfeind[2], T. Dollinger[2], K. Richter[2] and J. Nitta[1]*

[1]*Graduate school of Engineering, Tohoku University, 6-6-02 Aramaki-Aza Aoba, Aoba-ku, Sendai 980-8579, Japan*

[2]*Institut für Theoretische Physik, Universität Regensburg, 93040 Regensburg, Germany*

*Correspondence to: nitta@material.tohoku.ac.jp

**Spin-orbit interaction (SOI) plays a crucial role in device fields from physics to applications, such as Majorana Fermions, topological insulators, quantum information and spintronics to name only a few. Two types of SOIs, the Rashba SOI $\alpha$ and the Dresselhaus SOI $\beta$, act on spins as effective magnetic fields with different directions. Although spin angular momentum is not a good quantum number in spin-orbit coupled systems, spin SU(2) symmetry is preserved and the so-called persistent spin helix state is realized when $\alpha$ is equal to $\beta$. Existing methods to evaluate $\alpha/\beta$ require fitting analyses that often include an ambiguity in the parameters used. Here, we experimentally demonstrate simple and fitting parameter-free techniques to determine $\alpha/\beta$ and to deduce the absolute values of $\alpha$ and $\beta$ by detecting the effective magnetic field direction and the strength induced by the two SOIs. Moreover, we observe the spin SU(2) symmetry by gate tuning.**

As an essential feature of spin-orbit interaction (SOI), the spin of an electron moving in an electric field experiences, in its rest frame, an effective magnetic field that couples to the spin's magnetic moment, even in the absence of any external magnetic field. So, the SOI is ubiquitous and appears in many different areas from physics to applications[1-3]. The SOI can be engineered by material design or by tuning an electric field[4,5] to explore new spin-related phenomena and spin functionalities. In semiconductor-based spintronics, for example, this effective magnetic field $B_{\text{eff}}$ can be used for generation[6], manipulation[7-12], and detection[13] of spins solely by electrical means. However, the SOI inevitably gives rise to spin relaxation[14], thus disrupting long spin coherence. This crucial problem can be overcome by employing two different types of SOIs; the Rashba[15] and the Dresselhaus[16] SOIs. Spin SU(2) symmetry is realized and spin relaxation is suppressed correspondingly when both SOIs are of equal strength, giving rise to the persistent spin helix (PSH) state[17-20]. As a result, precise evaluation and control of $\alpha/\beta$ ($\alpha$ and $\beta$ denote the strengths of Rashba and Dresselhaus SOIs, respectively) pave the way for future applications in spintronics and quantum information[21-23]. However, for transport measurements, simultaneous evaluation of Rashba and Dresselhaus SOIs usually involves large ambiguities due to model-dependent fitting[24], and in optical measurements[25,26], fitting analyses are also required. Consequently, there has been the quest for a simple, robust and reliable technology for evaluating $\alpha/\beta$ for a long time. Here, we demonstrate a fitting-free determination of in-situ $\alpha/\beta$ in transport measurements under the coexistence of Rashba and Dresselhaus SOIs. In narrow wires, application of an in-plane magnetic field $B_{\text{in}}$ modulates the amplitude of the magneto-conductance, i.e. weak localization (WL) through additional dephasing arising from spin-induced time reversal symmetry breaking[27]. Theory predicts the maximum amplitude of WL when $B_{\text{in}}$ and $B_{\text{eff}}$ are parallel[28]. This enables us to evaluate $\alpha/\beta$ from the anisotropic WL amplitude, without **relying on any fitting. By applying the proposed concept to InGaAs based wire structures, we realize the gate controlled PSH state and find a novel finger print of the PSH state. In addition to the evaluation of the ratio $\alpha/\beta$, the absolute values of $\alpha$ and $\beta$ can be deduced from the minimum WL amplitude when $B_{in}$ and $B_{eff}$ have equal strengths. Our demonstration of proposed metrological concept[28] can be applied to a wide range of materials, since the direction and the strength of the effective magnetic fields in a confined structure are directly related to $\alpha$ and $\beta$. As a proof of concept we provide measurements in InGaAs wires. The precise determination of $\alpha$ and $\beta$ is indispensable to explore new spin-related phenomena and functionalities.**

According to the proposed metrological concept[28], Evaluation of $\alpha/\beta$ is achieved by detecting $\theta_{\text{eff}}$ of $\mathbf{B}_{\text{eff}} = \mathbf{B}_R + \mathbf{B}_D$ direction, where $B_R$ and $B_D$ are the effective magnetic fields induced by the Rashba and linear Dresselhaus SOIs. For a quantum well grown in the [001] direction of III–V zinc-blende semiconductor heterostructures, The corresponding Hamiltonians are given by

$$H_R = \frac{\alpha}{\hbar}(\sigma_x p_y - \sigma_y p_x), \qquad (1)$$

$$H_D = \frac{\beta}{\hbar}(\sigma_x p_x - \sigma_y p_y), \qquad (2)$$

where $\alpha$ and $\beta$ are coefficients of the Rashba and Dresselhaus SOIs, which determine the strengths of $B_R$ and $B_D$, respectively. In equations (1,2) $\sigma_i$ and $p_i$ (i = x, y) are the Pauli spin matrices and the electron momentum. The x and y directions are parallel to [100] and [010], respectively. Let us assume a one dimensional (1D) wire along [100], since electron momentum is restricted to ±p along the wire direction. The direction of $B_{\text{eff}}$ acting on electron spins is fixed as uniaxial[29,30]. Owing to the perpendicular orientation of $B_R$ and $B_D$, the relationship between the SOI parameters and the effective field orientation $\theta_{\text{eff}}$ is simply given by

$$\frac{\alpha}{\beta} = -\tan\theta_{\text{eff}}. \qquad (3)$$

It should be noted that the cubic Dresselhaus SOI vanishes along the [100] axis. Hence, by the detection of $\theta_{\text{eff}}$ in the 1D wire, direct evaluation of the $\alpha/\beta$ ratio is possible.

The central concept to determine $\theta_{\text{eff}}$ through transport measurements is schematically shown in Fig. 1. For a narrow wire with $W \ll L_{\text{so}}$ (W is the wire width and $L_{\text{so}}$ is the spin precession length. $L_{\text{so}} = \hbar^2/2m^*\alpha$ or $\hbar^2/2m^*\beta$, where $m^*$ is an effective mass), theory predicts that the narrow wire can be treated as quasi-one dimensional, and the $B_{\text{eff}}$ direction becomes uniaxial (green arrows in Fig. 1)[28]. Since the randomization of the spin precession axes is suppressed due to the one dimensional confinement of electron momentum p, the spin relaxation length is enhanced in a narrow wire and the magneto-conductance exhibits weak localization (WL), predicted theoretically[29,30] and confirmed experimentally[31,32]. The application of a static in-plane magnetic field $B_{in}$ further induces the additional spin relaxation by breaking the uniaxial alignment of $B_{eff}$. The additional spin relaxation induces dephasing in the time-reversal related pairs of electron paths by mixing spin up and down phases i.e.

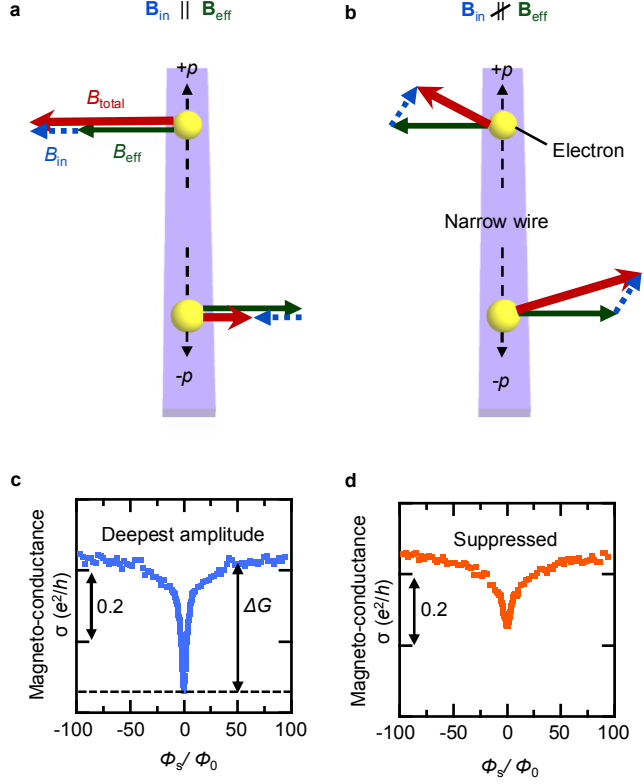

**Figure 1: WL anisotropy to the relative angle between $B_{eff}$ and $B_{in}$.** In the upper panels **a,b**, the directions of total magnetic field $\mathbf{B}_{total} = \mathbf{B}_{in} + \mathbf{B}_{eff}$ are shown for $\pm p$ momentum directions. In the lower panels **c,d**, calculated WLs are shown at $\alpha/\beta = 2/3$ in a [100] wire with $|\mathbf{B}_{in}| = 0.75$ T for $\mathbf{B}_{in} \parallel \mathbf{B}_{eff}$ (**c**: $\theta_{in} = 146°$) and $\mathbf{B}_{in} \nparallel \mathbf{B}_{eff}$ (**d**: $\theta_{in} = 25°$), where $\theta_{in}$ is the angle of in-plane magnetic field defined with respect to the [100] direction. The magneto-conductance is plotted against the magnetic flux in a unit cell of the numerical grid $\Phi_s = a^2|\mathbf{B}_p|$ ($a$ is the lattice constant and $B_p$ is the perpendicular magnetic field) in units of the magnetic flux quantum $\Phi_0 = h/e$. **a,c** When $\mathbf{B}_{in} \parallel \mathbf{B}_{eff}$, the total magnetic field $B_{total}$ along the $\pm p$ directions remains uniaxial showing the deepest WL amplitude. **b,d**, When $\mathbf{B}_{in} \nparallel \mathbf{B}_{eff}$, $B_{total}$ is randomized for $\pm p$ electron directions, suppressing the WL amplitude.

spin-induced time-reversal symmetry breaking[27]. This dephasing leads to weaker quantum interference, resulting in suppression of the WL amplitudes. When $B_{in}$ is applied in parallel to $B_{eff}$ (Fig. 1a), the total magnetic field $\mathbf{B}_{total} = \mathbf{B}_{in} + \mathbf{B}_{eff}$ for $\pm p$ directions is still uniaxial, preserving the long spin relaxation length. However, when $B_{in}$ is no longer parallel to $B_{eff}$ (Fig. 1b), the direction of $B_{total}$ is different for $\pm p$ momenta, enhancing the additional spin relaxation. Since the electron dephasing induced by the spin relaxation process decreases the amplitude of WL, the WL amplitude becomes anisotropic as a function of the relative angle between $B_{in}$ and $B_{eff}$ (Fig. 1c and 1d). This concept enables us to determine $\theta_{eff}$ by tilting the direction of $B_{in}$, and therefore, to determine the $\alpha/\beta$ ratio by using the relation of equation (3) without any fittings.

In order to confirm the concept presented above, we conduct numerical calculations of the magneto-conductance of wires with different $\alpha/\beta$ values including the persistent spin helix (PSH) configuration. In this calculation, we include also results for wires oriented along the [110] and [1-10] directions and compare our results with the data for the wire in [100] direction, since wires along [110] and [1-10] provide parallel and anti-parallel configurations between $B_R$ and $B_D$, respectively. We conduct the numerical calculations based on the recursive Green's function approach[28] (for details, see supplementary information). We theoretically calculate the magneto-conductance of a narrow wire by changing the perpendicular magnetic field $B_p$ under the fixed in-plane magnetic field $B_{in}$ whose angle $\theta_{in}$ is defined with respect to the [100] direction. Three different wire directions are examined along the [100], [110] and [1-10] axes, and are referred to as [100], [110] and [1-10] wire, respectively. It should be noted that the wire width is set to 20 nm, which does not give rise to a purely 1D confinement, but is smaller than the spin precession length. Figures 1c and 1d show the calculated results for the [100] wire at $\theta_{in}$ = 25º and 146º ($B_{in}$ = 0.75 T), exhibiting clear WL signals with different conductance amplitude. We define the amplitude of WL as $\Delta G$ (Fig. 1c) and produce a polar-plot as a function of $\theta_{in}$ for different $\alpha/\beta$ ratio and wire directions (Figs. 2a – 2c). The WL amplitude shows two-fold symmetries for all wire directions. Furthermore, its symmetry axis depends on the crystal direction. Such dependences are understood in terms of the relative angle between $B_{in}$ and $B_{eff}$.

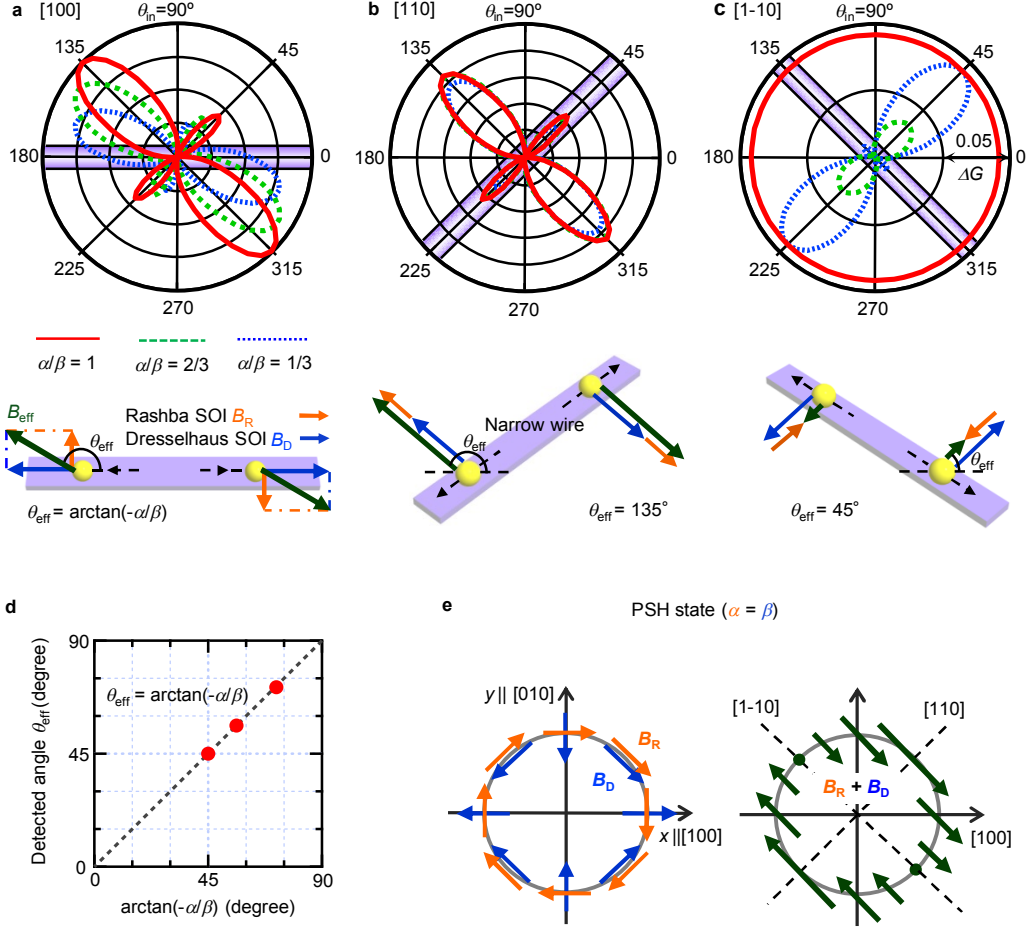

**Figure 2: Numerical results of conductance in quasi-one dimensional quantum wires.** The calculation is performed for $W \ll L_{so}$ without cubic Dresselhaus SOI. We fix the ratio $\alpha/\beta$ at 1/3 (blue dotted line), 2/3 (green dashed line) and 1 (red solid line). **a,b,c**, polar-plots of WL amplitudes as a function of $\theta_{in}$ along [100], [110] and [1-10] directions. Schematic insets indicate the angles of effective magnetic field $\theta_{eff}$ for each wire orientation. **d**. Detected $\theta_{eff}$ was perfectly fitted in accordance with relation $\theta_{eff} = \arctan(-\alpha/\beta)$. **e**. Orientation of $B_{eff}$ in momentum space under the persistent spin helix state (PSH state: $\alpha = \beta$). The arrows indicate the direction of the effective magnetic field $B_{eff}$ describing spin-orbit coupling. The length of the shown arrows indicates the strength of $B_{eff}$. In the left panel, Rashba and Dresselhaus effective magnetic fields ($\mathbf{B}_R$ and $\mathbf{B}_D$ respectively) are shown. In the right panel, total effective magnetic field $\mathbf{B}_{eff} = \mathbf{B}_R + \mathbf{B}_D$ is shown. The PSH state induces uniaxial alignment of $B_{eff}$.

Since the $\mathbf{B}_{in} \parallel \mathbf{B}_{eff}$ configuration does not enhance the spin relaxation, the angle of the maximum WL amplitude corresponds to the $B_{eff}$ direction. On the other hand, the $\mathbf{B}_{in} \nparallel \mathbf{B}_{eff}$ configuration induces spin relaxation resulting in a reduced WL amplitude. For [110] and [1-10] wires, since $B_{eff}$ appears always along 135º and 45º directions, respectively, the WL amplitude becomes maximal at $\theta_{in}$ = 135º and at 45º, and is thus unaffected by the different $\alpha/\beta$ ratio (Fig. 2b and 2c). For the [100] wire, however, the $B_{eff}$ direction changes with the $\alpha/\beta$ ratio, resulting in the angular shift of the maximum WL amplitude (Fig. 2a). As shown in Fig. 2d, the angle of the maximum WL amplitude ($\theta_{eff}$) is perfectly aligned to equation (3). In the PSH state, i.e., at $\alpha/\beta$ = 1 (red plot in Fig. 2a – c), while the maximum WL peak still appears at $\theta_{in}$ = 135º for the [110] wire, interestingly, the anisotropy of the WL amplitude is quenched for the [1-10] wire. This is due to the mutual compensation of $B_R$ and $B_D$ (Fig. 2e). This numerically observed quench of the WL anisotropy is a finger print for the PSH states.

In order to utilize this concept in a real device, we fabricate parallel wire structures based on InGaAs, as shown in Fig. 3 (more information on the device structure and fabrication is found in Methods). An n-$In_{0.52}Al_{0.48}As$ / 10 nm $In_{0.7}Ga_{0.3}As$ / n-$In_{0.52}Al_{0.48}As$ quantum well is designed to minimize the Rashba SOI to values of the same order as the Dresselhaus SOI by Si doping in both $In_{0.52}Al_{0.48}As$ barriers. We prepare parallel wires along the [100], [110], and [1-10] directions in series (Fig. 3) and cover the entire structure with an $Al_2O_3$ / Cr / Au top gate electrode to control $\alpha$ through the gate voltage ($V_g$). The wire width of 750 nm is set to be shorter than the spin precession length $L_{so} = \hbar^2 / 2m^*\alpha \approx 1.79$ μm at carrier density $N_s$ = $2.0 \times 10^{12}$ cm$^{-2}$ in the InGaAs 2DEG (two dimensional electron gas) channel. $N_s$ and electron mobility $\mu$ at $V_g$ = 0 V are $2.44 \times 10^{12}$ cm$^{-2}$ and 6.68 m$^2$/Vs, respectively. The magneto-conductance measurement is simultaneously performed for three wire directions by sweeping $B_p$ under the constant $B_{in}$ and $V_g$ at $T$ = 1.7 K.

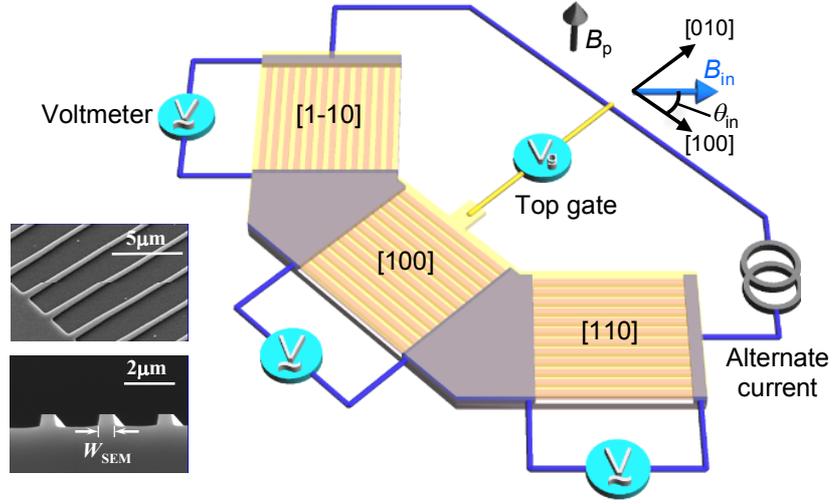

**Figure 3: Schematic illustration of measurement configuration.**
The shown setup allows us to simultaneously perform conductance measurements in all of the three considered wire orientations along the [100], [110] and [1-10] axes, respectively. The conductance for each wire orientation is an average over multiple wires in order to minimize signatures from the universal conductance corrections. The scanning electron microscope (SEM) images at the lower left are the example of InGaAs wire structure. Each image shows top view (upper picture) and side view (lower picture), respectively.

Figure 4a shows the magneto-conductance for the [100] wire at $\theta_{in}$ = 24º and 142º with $|\mathbf{B}_{in}|$ = 1.0 T and $V_g$ = -5.0 V. The restricted electron momentum in a narrow wire suppresses the spin relaxation, resulting in WL signals. In addition, modulation of the WL amplitude $\Delta\sigma$ between $\theta_{in}$ = 24º and 142º indicates the anisotropic dephasing due to additional spin relaxation. In order to reveal the effect of $B_{in}$ to the WL amplitude, we measure the magneto-conductance between 0º ≤ $\theta_{in}$ ≤ 180º (Fig. 4b) and extract $\Delta\sigma$ as a function of $\theta_{in}$ (Fig. 4c). The WL amplitude is continuously changed as shown in Fig. 4b and the oscillatory behavior of $\Delta\sigma$ in Fig. 4c indicates transitions between suppression and enhancement of electron dephasing, i.e. the modulation of spin relaxation. In order to compare with the theoretical prediction (Fig. 2a-d), we include a polar plot of $\Delta\sigma'$ (subtracted $\Delta\sigma$ from its minimum) as a function of $\theta_{in}$ (Fig. 4d).

Due to the spatial symmetry of the configuration, we safely extend the result to $180° \leq \theta_{in} \leq 360°$. This plot represents how additional spin relaxation is induced by the relative angle between $B_{in}$ and $B_{eff}$. Since $\Delta\sigma'$ shows a two-fold symmetry which is perfectly consistent to the theoretical prediction, $\theta_{in}$ for the maximum $\Delta\sigma'$ corresponds to $\theta_{eff}$. The second minor maximum WL amplitude around the angle of 50° is also well reproduced in the experiment, which originates from the preservation of the orthogonal symmetry class. According to the theoretical study, in the case of $\mathbf{B}_{in} \perp \mathbf{B}_{eff}$, although there is no spin rotation symmetry, the spin rotated version of the Hamiltonian $\widetilde{H}_{Q1D}$ is given by $\hat{C}^{-1}\widetilde{H}_{Q1D}\hat{C} = \widetilde{H}_{Q1D}$, which belongs to the orthogonal symmetry class[33] where $H_{Q1D}$ is single particle Hamiltonian of disordered quantum wire (The definition of $H_{Q1D}$ is shown in the supplementary information.) and $\hat{C}$ is the operator of complex conjugation. This leads to the recovery of WL amplitude and appearance of minor peaks. Mentioned symmetry class describes the system because of a fixed $B_{eff}$. This is only true if the contributions of the momentum perpendicular to the wire are negligible. Hence, appearance of these minor peaks is strong evidence for dimensional control of the electron spin and the validity of the $\alpha/\beta$ evaluation technique in our measurement setup. Also, to confirm the dependence of our method on the strength of $B_{in}$, we include additional polar plots for changed values of $B_{in}$ from 0.2 to 1.0T, while leaving the other parameters unaltered (Fig. 4e). According to the theory[28], detection technique of $\alpha/\beta$ cannot be correctly applicable when $|\mathbf{B}_{in}| \gg |\mathbf{B}_{eff}|$ since $B_{total}$ gets strongly aligned in the direction of $B_{in}$ and electron dephasing is suppressed for any $\theta_{in}$. This potentially changes the major peak angle. However, no shift of the major peak is observed at different $B_{in}$, which guarantees high accuracy in the direct detection of $\alpha/\beta$. (The $|\mathbf{B}_{in}|$ dependences for other wire directions are discussed in the supplementary information.)

To further confirm the validity of this concept, we plot $\Delta\sigma'$ for all considered wire directions with different $V_g$ (Fig. 5). For the [110] and [1-10] wires, the two-fold symmetry of $\Delta\sigma$ is observed. The $\Delta\sigma'$ maxima correspond to 135º and 45º, respectively, consistent with thepredicted angle of $B_{eff}$ calculated in Fig. 2b and 2c. Since $B_{eff}$ is always oriented perpendicular to the [110] and [1-10] wire direction (purple wires in Figs. 5b and 5c) for different $V_g$, $\theta_{in}$ at the $\Delta\sigma'$ maximum does not change according to the $\alpha/\beta$ ratio. From these considerations, we directly probe the precise direction of $B_{eff}$ by measuring the WL maximum. Moreover, we can extract the $\alpha/\beta$ ratio from the anisotropy of [100] wire. As shown in Fig. 5a, the $\Delta\sigma'$ maximum is shifted towards 135º by decreasing $V_g$ from 0V to -9V, showing the $\alpha/\beta$ modulation through the gate. Since $\alpha$ is increased with reducing $N_s$ (2.44×10$^{12}$ cm$^{-2}$ at $V_g = 0$V to 2.01×10$^{12}$ cm$^{-2}$ at $V_g = -9$V) and is smaller than $\beta$ at $V_g = 0$V, the $\alpha/\beta$ ratio is approaching unity at $V_g = -9$V. By using equation (3), we determine the $\alpha/\beta$ ratio as, $\alpha/\beta = -\tan(157º \pm 1º) = 0.42 \pm 0.02$ at $V_g = 0$V, $\alpha/\beta = -\tan(142º \pm 1º) = 0.78 \pm 0.03$ at $V_g = -5$V, and $\alpha/\beta = -\tan(133º \pm 1º) = 1.07 \pm 0.04$ at $V_g = -9$V.

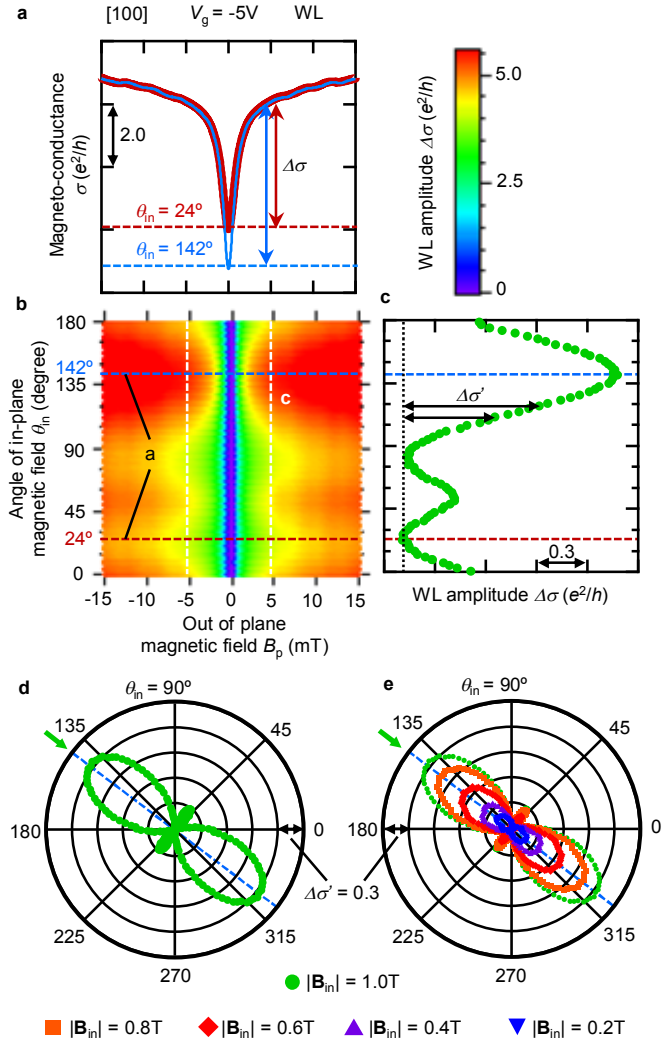

**Figure 4: WL anisotropy in [100] direction at $V_g$ = -5V. a**. Measured magneto-conductance traces for $\theta_{in}$ = 24º, 142º. $\Delta\sigma$ is the WL amplitude extracted at $|\mathbf{B}_p|$ = 5 mT. **b**. Color plot of WL signals measured between 0º ≤ $\theta_{in}$ ≤ 180º. The zero point is chosen as the minimum of the magneto-conductance ($\sigma$ at $|\mathbf{B}_p|$ = 0 mT). **c**. WL amplitude $\Delta\sigma$ depending on different $\theta_{in}$. $\Delta\sigma'$ is the magneto-conductance difference from minimum $\Delta\sigma$. **d**. Polar plot of $\Delta\sigma'$ between 0º ≤ $\theta_{in}$ ≤ 360º in $|\mathbf{B}_{in}|$ = 1.0 T. We extended the result of 0º ≤ $\theta_{in}$ ≤ 180º to 180º ≤ $\theta_{in}$ ≤ 360º considering the spatial symmetry. **e**. Polar plot adding different $|\mathbf{B}_{in}|$ = 0.2 to 1 T.

The condition of PSH state is satisfied by the $\alpha/\beta$ ratio at $V_g = -9$V. As shown in Fig. 2e, the PSH state induces the uniaxial alignment of $B_{eff}$ along the [1-10] direction with modulated strength. Especially, $B_{eff}$ vanishes in the [1-10] axis due to the mutual cancellation of $B_R$ and $B_D$. As shown in Fig. 5c, for [1-10] wires, the anisotropy of $\Delta\sigma'$ exhibits the two-fold symmetry along the [110] axis for $V_g = 0$V and -5V due to the presence of $B_{eff}$. In contrast to the other gate voltage conditions, the anisotropy is quenched at $V_g = -9$V. This is caused by the dominant strength of $B_{in}$ over $B_{eff}$, which gives rise to a homogeneous dephasing rate regardless of the $\theta_{in}$ direction and the shape of polar-plot becomes a circle when we add the offset of the magneto-conductance ($V_g = -9$V in Fig. 5c). This quench of the WL anisotropy agrees with the numerical result and it can be regarded as the characteristic finger print to support the realization of PSH state revealed by the present method.

We confirm that the WL amplitude shows a minimum when $|\mathbf{B}_{in}|$ is close to $|\mathbf{B}_{eff}|$[28,33]. By detecting $|\mathbf{B}_{eff}|$ from the $|\mathbf{B}_{in}|$ dependence of WL amplitude, we can estimate the absolute values of $\alpha$ and $\beta$ to an accuracy of up to a factor of two. The numerical calculation shows minima of the WL amplitude if the ratio $|\mathbf{B}_{in}|/|\mathbf{B}_{eff}|$ is close to 1 (Fig. 6a). This calculated result is also consistent with our analytical toy model, describing the spin interference term of two one

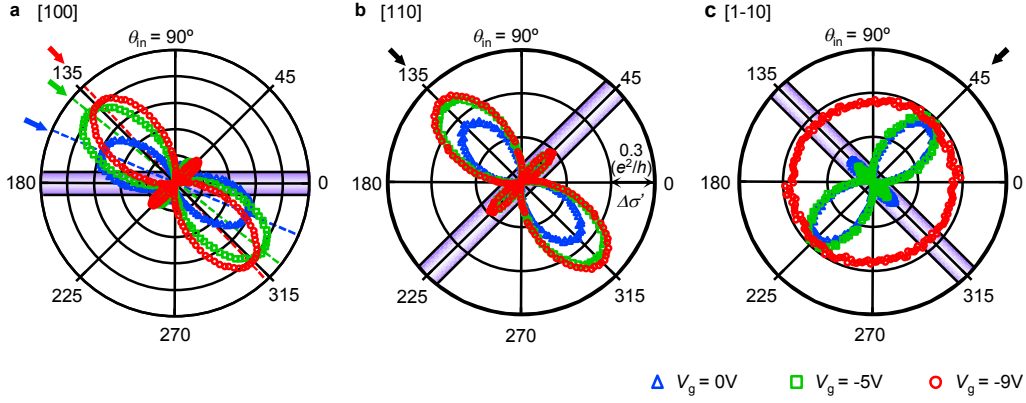

**Figure 5: Polar-plots of WL amplitudes as a function of $\theta_{in}$. a,b,c**, The polar-plots in [100], [110] and [1-10] direction at $V_g = 0$V (Blue triangle), $V_g = -5$V (Green rectangle) and $V_g = -9$V (Red circle). Respective polar-plots were made in the same way as Fig. 4d. Plots are normalized fixing the minimum WL amplitude as zero except for $V_g = -9$V in [1-10] direction.

dimensional time-reversal symmetry related paths (see supplementary information for details). We experimentally observe the $|\mathbf{B}_{in}|$ dependence of 'difference of the WL amplitudes $G' = \Delta\sigma(\theta \neq 0º) - \Delta\sigma(\theta = 0º)$' at $V_g = -5$V (Carrier density $N_s = 2.16 \times 10^{-12}$ cm$^{-2}$), where $\Delta\sigma$ is a WL amplitude and $\theta$ is an angle between $\mathbf{B}_{in}$ and $\mathbf{B}_{eff}$ (Fig. 6b-d). We used the WL amplitude at $\theta = 0º$ as reference value for the remaining WL amplitudes. The characteristic line shapes of the numerical (Fig. 6a) and the experimental (Fig. 6b-d) datasets exhibit remarkable resemblance. Moreover, when we compare the traces in each wire, $|\mathbf{B}_{in}| \approx |\mathbf{B}_{eff}|$ at dip position is the largest in [110] wire and smallest in [1-10] wire. This result agrees with the relation between $|\mathbf{B}_{eff}|$ and $|\mathbf{B}_R|$, $|\mathbf{B}_D|$ in each crystal direction: $|\mathbf{B}_{eff}| = |\mathbf{B}_R| + |\mathbf{B}_D|$ ([110] wire) $> |\mathbf{B}_{eff}| = \sqrt{|\mathbf{B}_R|^2 + |\mathbf{B}_D|^2}$ ([100] wire) $> |\mathbf{B}_{eff}| = |\mathbf{B}_R| - |\mathbf{B}_D|$ ([1-10] wire). In particular along the [100] axis, cubic Dresselhaus SOI is negligible, hence $\alpha$ and $\beta$ can be deduced by using the estimated $|\mathbf{B}_{eff}|$ and the $\alpha/\beta$ ratio obtained from WL anisotropy measurements in the [100] wire. We estimate $\alpha$ and $\beta$ approximately as $\alpha \approx 2.0 \times 10^{-13}$ eV·m, $\beta \approx 3.7 \times 10^{-13}$ eV·m by using the evaluated $\alpha/\beta = -\tan(\theta_{eff} \approx 152º) \approx 0.53$ and $|\mathbf{B}_{eff}| \approx 1.55$ T. The small value of $\alpha$ of the order of $10^{-13}$ eV·m is reasonable, since our heterostructure design correspond to a symmetric quantum well (QW).

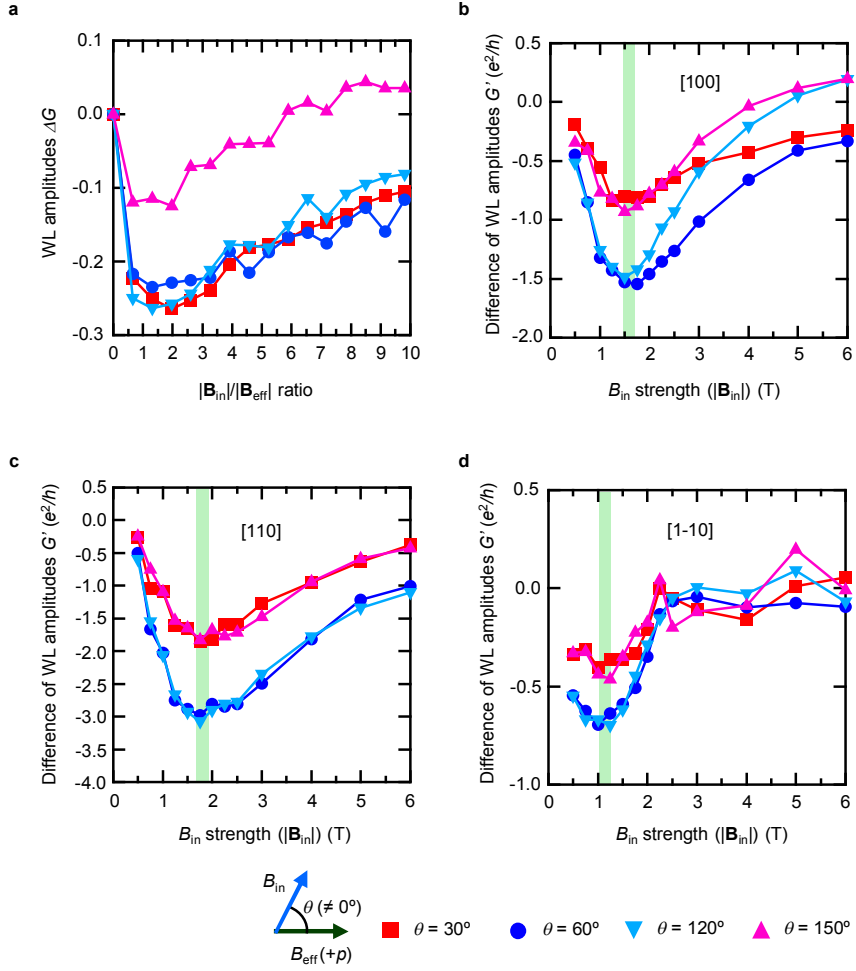

**Figure 6: WL amplitude as a function of the $|B_{in}|/|B_{eff}|$ ratio.** Figure 6a shows the numerical calculation result. The experimental results are shown in Fig. 6b-d. **a.** The calculation is performed for $W \ll L_{so}$ without cubic Dresselhaus SOI in a [100] quasi-one dimensional wire. We show the dependence of the WL amplitude $\Delta G$ as a function of the ratio $|B_{in}|/|B_{eff}|$. To regard $\Delta G(|B_{in}| = 0)$ the basis 0, the offset $- \Delta G(|B_{in}| = 0)$ is added. The $\theta$-dependence (the angle between $B_{in}$ and $B_{eff}$) is shown for $\theta = 30°$ (red rectangle), $\theta = 60°$ (blue circle), $\theta = 120°$ (light blue downward triangle) and $\theta = 150°$ (light red upward triangle). We observe clear cut minima of the WL conductance if the ratio $|B_{in}|/|B_{eff}|$ is close to 1. **b,c,d,** The experimentally obtained WL amplitudes $G' = \Delta\sigma(\theta \neq 0°) - \Delta\sigma(\theta = 0°)$ at $V_g = -5V$ (carrier density $N_s = 2.16 \times 10^{12}$ cm$^{-2}$) are plotted for the [100], [110] and [1-10] wires, respectively. Each angle $\theta (\neq 0°)$ is shown as $\theta = 30°$ (red rectangle), $\theta = 60°$ (blue circle), $\theta = 120°$ (light blue downward triangle) and $\theta = 150°$ (light red upward triangle).

The magnitude obtained for $\beta$ well agrees with reported values in InGaAs QW: $\beta \approx 3.5\times10^{-13}$ eV·m (Ref. 34).

In conclusion, we have demonstrated a novel metrological concept and technique to evaluate the $\alpha/\beta$ ratio, which can be applied in different material systems. The present concept enables us to determine the effective magnetic field direction, associated with the SOIs, to a high accuracy and furthermore allows for the evaluation of the $\alpha/\beta$ ratio, including the correct sign, without fitting. We have also realized the gate controlled PSH state and have found a new finger print indicating the realization of the PSH state. This method can determine the sign of the $\alpha/\beta$ ratio, so it is possible to detect an inverted PHS (i-PHS) state with $-\alpha/\beta$ by tuning the gate the voltage. For example, the gate controlled transition between the PHS and i-PHS states provides possibilities for future spintronic devices requiring large spin coherence times[23]. Furthermore, we have proposed a new technique to estimate the absolute values of $\alpha$ and $\beta$ by detecting the strength of $B_{\text{eff}}$ without fitting. These results pave the way to explore new spin-related phenomena and to facilitate the design of future spintronic and quantum information devices based on both Rashba and Dresselhaus SOIs.

**Methods**

For our measurement, we used following heterostructure: $In_{0.52}Al_{0.48}As$ (200 nm, buffer layer)/$In_{0.52}Al_{0.48}As$ (6 nm, carrier supply layer; Si doping concentration of $2\times10^{18}$ cm$^{-3}$)/$In_{0.52}Al_{0.48}As$ (6 nm, spacer layer)/$In_{0.7}Ga_{0.3}As$ (10 nm, quantum well)/$In_{0.52}Al_{0.48}As$ (6 nm, spacer layer)/$In_{0.52}Al_{0.48}As$ (6 nm, carrier supply layer; Si doping concentration of $2\times10^{18}$ cm$^{-3}$)/$In_{0.52}Al_{0.48}As$ (25 nm, cap layer). We made a symmetric quantum well (QW) by doping on both sides of the QW. Owing to the symmetric shape of QW, $\alpha$ was adjusted to a small value to match $\beta$. Furthermore, to suppress an influence of cubic Dresselhaus SOI, we used strained QW ($In_{0.7}Ga_{0.3}As$) with additional $\beta$: strain induced Dresselhaus SOI $\beta_{\text{strain}}$[35,36]. These structures were epitaxially grown on a (001) InP substrate by metal organic chemical vapour deposition. The epitaxial wafer was processed into narrow wire structures using e-beam lithography and reactive-ion etching. Structural parameters of the wire are as follows: wire length '$L$' = 200 μm, effective width '$W$' = 750 nm and number of parallel wires '$N$' = 100 to average out universal conductance fluctuations. To apply $V_g$, we deposited 200 nm $Al_2O_3$ and 150 nm Cr/Au as an insulator and a top gate by atomic layer deposition and e-beam deposition. The measurement was carried out by AC lock-in technique for all wire sets at 1.7K. Sheet carrier density $N_s$ was deduced from the fast Fourier transformation of the Shubnikov-de Haas (SdH) oscillations. The perpendicular magnetic field is swept in ± 15 mT and fixed in-plane magnetic field is used ranging from 0.2 to 1.0 T.

**Acknowledgements**

We acknowledge support from the Strategic Japanese-German Joint Research Program. K.R. thanks the DFG for support within Research Unit FOR 1483. T.D. acknowledges support by the DFG within the research project SFB 689. This work was financially supported by Grants-in-Aid from the Japan Society for the Promotion of Science (JSPS) No. 22226001.

**Author Contributions**

A.S., S.N. and Y.N. performed device fabrication and measurements. T.B., T.D. and K.R. performed the numerical calculations. A.S. and M.K. wrote the main part of manuscript. T.D. and K.R. wrote the theoretical part. All authors discussed the results and worked on the manuscript at all stages. M.K., K.R. and J.N. planned the project. J.N. directed the research.

**Competing Financial Interests**

The authors declare no competing financial interests.

## Supplementary Notes

### I. Theoretical approach using recursive Green's function

In our numerical approach we use a finite difference scheme to represent the Hamiltonians $H_R$ and $H_D$ of the Rashba and the linear Dresselhaus SOI

$$H_R = \frac{\alpha}{\hbar}(\sigma_x p_y - \sigma_y p_x), \quad (S1)$$

$$H_D = \frac{\beta}{\hbar}[(\sigma_x \cos 2\varphi - \sigma_y \sin 2\varphi)p_x - (\sigma_x \sin 2\varphi + \sigma_y \cos 2\varphi)p_y], \quad (S2)$$

on a quadratic lattice. Here $\sigma_i$ are the Pauli spin matrices and $p_i$ (i = x, y) is the in-plane momentum. The x, y directions are chosen parallel and perpendicular with the narrow wire direction, respectively. $\varphi$ denotes the angle between x direction and [100] direction of the zinc blend crystal.

The relation between the SOI parameters ($\alpha$, $\beta$) and the direction of effective magnetic field ($B_{eff}$) $\theta_{eff}$ can be expressed by following equation[28].

$$-\frac{\alpha \cos\varphi + \beta \sin\varphi}{\alpha \sin\varphi + \beta \cos\varphi} = \tan\theta_{eff}. \quad (S3)$$

The angle $\theta_{eff}$ is measured with respect to the [100] direction. Especially in [100] direction ($\varphi = 0°$), equation (S3) can be simplified to the expression:

$$\varphi = 0° \, // \, [100]: \theta_{eff} = \arctan\left(-\frac{\alpha}{\beta}\right). \quad (S4)$$

This establishes a connection between the $\alpha/\beta$ ratio and the angle $\theta_{eff}$, which can be detected. This allows for a quantitative estimation of $\alpha/\beta$. As stated in the main text, measurement of weak localization (WL) in quasi-one dimensional quantum wire with in-plane magnetic field ($B_{in}$) makes it possible to detect $\theta_{eff}$, which leads to the evaluation of $\alpha/\beta$.

In this paper, to evaluate $\alpha/\beta$ for different gate voltages ($V_g$) in the vicinity of the persistent spin helix (PSH) state, we conduct numerical calculations for different $\alpha/\beta$ ratios. In addition to the [100] wire orientation, we model also [110] and [1-10] wires for comparison.

The corresponding characteristic values of $\theta_{\text{eff}}$ are determined by equation (S3):

$$\varphi = 45° \,//\, [110]: \theta_{\text{eff}} = 135°, \quad (S5)$$

$$\varphi = 135° \,//\, [1-10]: \theta_{\text{eff}} = 45°. \quad (S6)$$

Using an optimized recursive Green's functions method[33], we calculated the total quantum transmission probability $T(E_F)$ at Fermi energy $E_F$, which yields the conductance in linear response within the Landauer approach: $G = G_0 T(E_F)$. For details, see Refs. 37 and 38. We use the following single particle Hamiltonian to model transport in a disordered quantum wire:

$$H_{\text{Q1D}} = \frac{p_x^2 + p_y^2}{2m^*} + U_{\text{conf}}(y) + U_{\text{dis}}(x,y) + \frac{\mu_B g}{2}\left[\mathbf{B}_{\text{eff}}(p_y=0) + \mathbf{B}_{\text{in}}\right]\cdot\boldsymbol{\sigma}. \quad (S7)$$

In equation (S7), the in plane magnetic field contributing to the Zeeman interaction is given by

$$\mathbf{B}_{\text{in}} = B_{\text{in}}\left[\cos(\theta_{\text{in}} - \varphi)\hat{e}_x + \sin(\theta_{\text{in}} - \varphi)\hat{e}_y\right]. \quad (S8)$$

where $\theta_{\text{in}}$ is the angle between the magnetic field and the [100] direction. The effective magnetic field of the spin-orbit interaction is given by

$$\mathbf{B}_{\text{eff}} = \frac{2}{\mu_B g \hbar}\left\{\left[\hat{e}_x \beta\cos 2\varphi - \hat{e}_y(\alpha + \beta\sin 2\varphi)\right]p_x + \left[\hat{e}_x(\alpha - \beta\sin 2\varphi) - \hat{e}_y \beta\cos 2\varphi\right]p_y\right\}.$$

$$(S9)$$

The product of above expression with the vector of the Pauli matrices is, up to constant factors, given by the superposition of $H_R$ and $H_D$. Moreover, $U_{\text{conf}}(y)$ in the equation (S7) represents the hard wall confining potential for the quantum wire. $U_{\text{dis}}(x,y)$ is the disorder potential, which is implemented as Anderson disorder, following a uniform random distribution. In a previous study[28], the occurrence of a maximum of the WL amplitude for $\mathbf{B}_{\text{in}} \parallel \mathbf{B}_{\text{eff}}$ is explained theoretically within the random matrix theory (RMT) framework[39]. According to this theory, the size of the WL amplitude depends on the symmetry class attributed to the system. In this calculation, since $B_{\text{eff}}$ is independent of $p_y$ and the direction of $B_{\text{eff}}$ is fixed, $B_{\text{eff}}(p_y = 0)$, the system possesses U(1) spin rotation symmetry, implying conservation of spin. With an additional magnetic field $B_{\text{in}}$, usually, the U(1) spin rotation symmetry is broken for a total magnetic fields $B_{\text{total}}$ which points into different directions, depending on the sign of $p_y$. Consequently, spin relaxation is enhanced and WL is suppressed. However, in the special case of $\mathbf{B}_{\text{in}} \parallel \mathbf{B}_{\text{eff}}$, the Hamiltonian can be written in block diagonal form:

$$H_{\text{Q1D}} = \begin{pmatrix} H_+ & 0 \\ 0 & H_- \end{pmatrix}, \quad \text{(S10)}$$

where

$$H_\pm = \frac{p_x^2 + p_y^2}{2m^*} + U_{\text{conf}}(y) + U_{\text{dis}}(x,y) \pm \frac{\mu_B g}{2} B_{\text{in}} \pm \frac{1}{\hbar} \kappa' p_x, \quad \text{(S11)}$$

with $\kappa' = \sqrt{\alpha^2 + \beta^2 + 2\alpha\beta \sin 2\varphi}$. Using $\widetilde{H}_\pm = U_{\text{eff},\pm}^{-1} H_\pm U_{\text{eff},\pm}$ and $U_{\text{eff},\pm} = \exp(\mp im^*\kappa'x/\hbar^2)$, in the case of parallel effective and in-plane magnetic field, the commutation relation $\left[\widetilde{H}_\pm, \hat{C}\right] = 0$ is fulfilled, where $\hat{C}$ is the operator of complex conjugation. This implies time reversal symmetry of $H_+$ and $H_-$. Therefore, when $\mathbf{B}_{\text{in}} \parallel \mathbf{B}_{\text{eff}}$, the system is described by the orthogonal symmetry class and possesses spin rotation symmetry. This leads to a suppression of spin relaxation and maximum WL amplitude.

In our numerical simulation, we used the following parameters. Phase coherence length $L_\varphi$ = 1500 nm, spin precession length $L_{\text{so}} = \hbar^2/2m^*\alpha$ = 360 nm, mean free path $L_{\text{el}}$ = 202 nm, wire width $W$ = 20 nm and $E_F$ = 19 eV·m. In Fig. 3, we present the results of the numerical calculation for the average conductance differences $\Delta G$ (weak localization amplitude) as a function of $\theta_{\text{in}}$ in [100], [110] and [1-10] orientation for $\alpha/\beta$ = 1/3, 2/3, 1. Major peaks indicate the largest WL amplitude where $\theta_{\text{in}}$ corresponds to $\theta_{\text{eff}}$. As anticipated, the major peak position ($\theta_{\text{eff}}$) is modulated in [100] direction, while it is fixed in [110] and [1-10] directions. Moreover, in the PSH state ($\alpha/\beta = 1$), not only the major peak position ($\theta_{\text{eff}}$) points at 135° (= arctan(−1)) but also the quench of WL anisotropy occurs in [1-10] direction, which means the annihilation of the same strengths of $\alpha$ and $\beta$ in [1-10] direction due to the antiparallel orientations.

## II. WL anisotropy depending on $B_{in}$ strength

According to the theory[28], our proposed detection technique of $\alpha/\beta$ cannot be correctly applicable when $|\mathbf{B}_{in}| \gg |\mathbf{B}_{eff}|$ since $B_{total}$ gets strongly aligned in the direction of $B_{in}$ and spin dephasing is equally suppressed for any $\theta_{in}$. In the present experiment, the applied in-plane field is used by constant value of $|\mathbf{B}_{in}| = 1$ T. To confirm the validity in our measurement, we measure the $|\mathbf{B}_{in}|$ dependence of WL anisotropy by using different strengths of $B_{in}$ (0.2 to 1 T) at different gate voltages $V_g$. The WL anisotropy in different crystal directions are plotted in Fig. S1a-i. The detected angles $\theta_{eff}$ are almost fixed in different $|\mathbf{B}_{in}|$ in all wire directions, which indicates the validity of the $\alpha/\beta$ evaluation in our measurement. However, for lower $|\mathbf{B}_{in}|$, a slight shift of peak position is not so clear compared to the detected $\theta_{eff}$ in higher $|\mathbf{B}_{in}|$ fields. This can be interpreted in the following way. For smaller $|\mathbf{B}_{in}|$, such as $|\mathbf{B}_{in}| = 0.2$ T, $B_{in}$ is not sufficiently strong to generate the anisotropy of the WL amplitude under variation of $\theta_{in}$, which potentially induces a slight ambiguity in the detection of the angle $\theta_{eff}$. However, $\theta_{eff}$ is fixed with sufficiently large in-plane field $|\mathbf{B}_{in}| > 0.4$ T. We confirm the validity of our measurements for different gate voltages by the same method.

For the PSH state in the [1-10] direction, our numerical result shows that the WL anisotropy is quenched as $B_{in}$ becomes large compared to an annihilated $B_{eff}$ ($|\mathbf{B}_R| - |\mathbf{B}_D|$). The dominant $B_{in}$ generates a homogeneous dephasing rate and WL amplitudes regardless of $\theta_{in}$. However, in our experiment, a slight anisotropy of WL remains when $B_{in} = 0.6$ T (Fig. S1i). We attribute this to the fact that $B_R$ and $B_D$ are not completely compensated by each other since $\alpha/\beta$ is evaluated as $1.07 \pm 0.04$ in the [100] wire, which represents a minor deviation from $\alpha/\beta = 1$. Also, the quench of the WL anisotropy at $B_{in} = 0.2$ T can be inferred from the fact that a magnetic field of 0.2 T is not large enough to produce WL anisotropy. The WL anisotropy of $B_{in} = 0.6$T at $V_g = -9$V in the [1-10] wire is still much smaller compared to that at other values of $V_g$ i.e. WL anisotropy at $B_{in}$ = 0.6 T at $V_g$ = -5V in Fig. S1f. Considering these points, the quench of the anisotropy in our measurement can be regarded as a finger print of the PSH state to support the result that we find $\alpha/\beta$ = -tan($\theta_{eff}$ = 133º) = 1.07 ± 0.04 in [100] the wire.

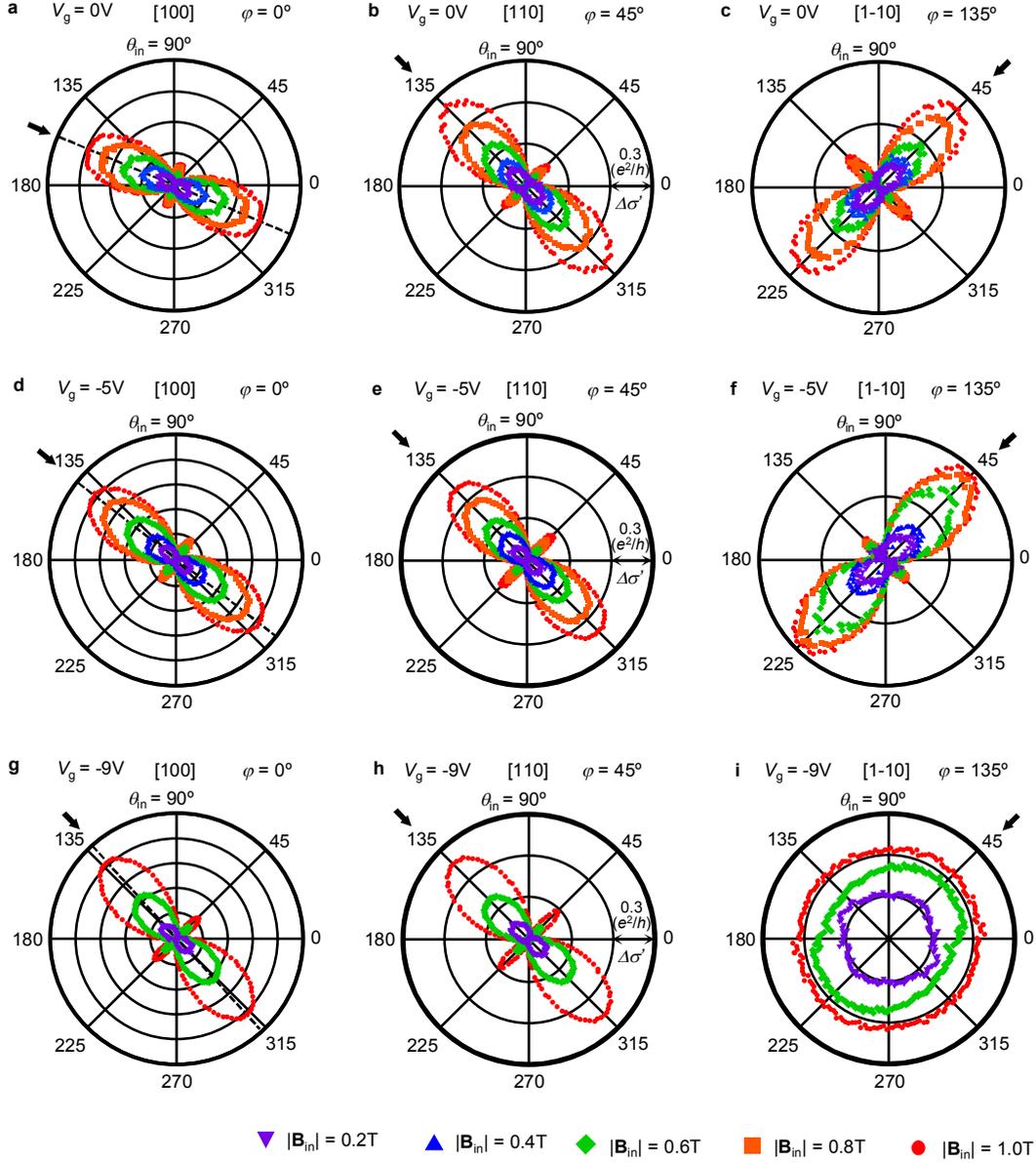

**Figure S1:** $B_{in}$ **dependence of the WL anisotropy. a-h,** WL anisotropy (WL amplitude $\Delta\sigma'$ as a function of $\theta_{in}$) depending on the strength of $B_{in}$ at $V_g$ = 0V, -5V and -9V in the [110], [100] and [1-10] wires respectively, except for $V_g$ = -9V in the [1-10] wire. We expose the sample to $|B_{in}|$ with magnitudes ranging from 0.2 to 1 T for each wire orientation. **i.** The WL anisotropy depending on the strength of $B_{in}$ at $V_g$ = -9V in [1-10] direction. The offset magneto-conductance is added to the measured values.

## III. Calculations for the evaluation of the absolute values of SOI parameters

It is suggested that the WL amplitude exhibits a minimum when $|\mathbf{B}_{in}|$ is very close to $|\mathbf{B}_{eff}|$ [28,33]. This is because spin relaxation is caused by randomization of $B_{total}$ which depends on an electron momentum. For example, when $|\mathbf{B}_{in}|$ is much larger than $|\mathbf{B}_{eff}|$, strong alignment of $B_{total}$ and $B_{in}$ suppresses spin relaxation, and when $|\mathbf{B}_{in}|$ is much smaller than $|\mathbf{B}_{eff}|$, the full WL amplitude is recovered. Therefore, the amplitude of WL is modulated by changing the ratio $|\mathbf{B}_{in}|/|\mathbf{B}_{eff}|$. From above simple arguments, we propose a technique to determine the absolute values of Rashba and Dresselhaus SOI parameters ($\alpha$ and $\beta$) as well as $|\mathbf{B}_{eff}|$.

This technique is based on the idea that spin relaxation is maximized when $B_{in}$ is very close to $B_{eff}$. Therefore, we first confirm this concept within a qualitative calculation based on a toy model. In the toy model, we calculate the probability of spin preservation for an electron spin which experiences spin precession during forward and backward propagation: denoted as clockwise (CW) and counterclockwise (CCW) paths. The schematic image of this model is shown in Fig. S2a. The vectors of $B_{total1}$ and $B_{total2}$ in the figure are defined by $\mathbf{B}_{total1} = \mathbf{B}_{eff}(+p) + \mathbf{B}_{in}$ and $\mathbf{B}_{total2} = \mathbf{B}_{eff}(-p) + \mathbf{B}_{in}$ respectively. Here, we define $\theta$ as the angle between $B_{in}$ and $B_{eff}$. $\theta_1$ and $\theta_2$ are angles of $\mathbf{B}_{total1}$ to $\mathbf{B}_{eff}(+p)$ and $\mathbf{B}_{total2}$ to $\mathbf{B}_{eff}(-p)$ (see Fig. S2a). To conduct the calculation, we define the spin rotation operators as

$$R_{\vec{m}}(\phi) = \mathbf{I}\cos(\phi/2) - i(\sigma \cdot \vec{m})\sin(\phi/2), \quad \text{(S11)}$$

$$R_{\vec{n}}(\phi) = \mathbf{I}\cos(\phi/2) - i(\sigma \cdot \vec{n})\sin(\phi/2), \quad \text{(S12)}$$

where $\mathbf{I}$ is a unit matrix and $\phi$ are spin precession angles around the axis of $B_{total1}$ and $B_{total2}$ respectively. $\vec{m} = (\cos\theta_1, \sin\theta_1, 0)$ and $\vec{n} = (-\cos\theta_2, \sin\theta_2, 0)$ are unit vectors of $B_{total1}$ and $B_{total2}$ respectively. The final spin states along CW and CCW paths are given by the following expressions:

$$|f_{CW}\rangle = R_{\bar{m}}(\phi_1)R_{\bar{n}}(\phi_2)R_{\bar{m}}(\phi_3)|i_{spin}\rangle, \quad (S13)$$

$$|f_{CCW}\rangle = R_{\bar{n}}(\phi_{1'})R_{\bar{m}}(\phi_{2'})R_{\bar{n}}(\phi_{3'})|i_{spin}\rangle. \quad (S14)$$

where $|i_{spin}\rangle$ is the initial spin state and we set $|i_{spin}\rangle = \frac{1}{2}\begin{pmatrix}1\\0\end{pmatrix}$ for convenience. The probability of spin preservation $P$ in the interference at the original point is described as

$$P = \left(|f_{CW}\rangle + |f_{CCW}\rangle\right)^{+}\left(\langle f_{CCW}| + \langle f_{CW}|\right). \quad (S15)$$

Since $\theta_1$, $\theta_2$, $B_{total1}$ and $B_{total2}$ are related to $B_{in}$, $B_{eff}$ and $\theta$, the spin preservation probability $P$ is regarded as a function of $B_{in}$, $B_{eff}$, and $\theta$. Here, we fix $|\mathbf{B}_{eff}| = 1$ and calculate the probability $P$ by varying $B_{in}$ for several different angles $\theta$. The calculated results are shown in Fig. S2b. When $\theta = 0°$, $180°$, the spin preservation probability is fixed at unity, which indicates that no additional spin relaxation is induced by $B_{in}$ regardless of the magnetic field strengths. This can be explained by the fact that spin relaxation is suppressed for uniaxial $B_{total}$. In contrast, for the case of other angles such as $\theta = 30°$, $60°$, $120°$, $150°$, the spin preservation probability changes as a function of the magnitude of $B_{in}$ and gets minimal when $|\mathbf{B}_{in}| \approx |\mathbf{B}_{eff}| = 1$, as expected. Thus, we confirm that the spin relaxation rate is maximized when the external magnetic field $B_{in}$ is comparable with $B_{eff}$ except for $\theta = 0°$, $180°$ ($\mathbf{B}_{in} \parallel \mathbf{B}_{eff}$).

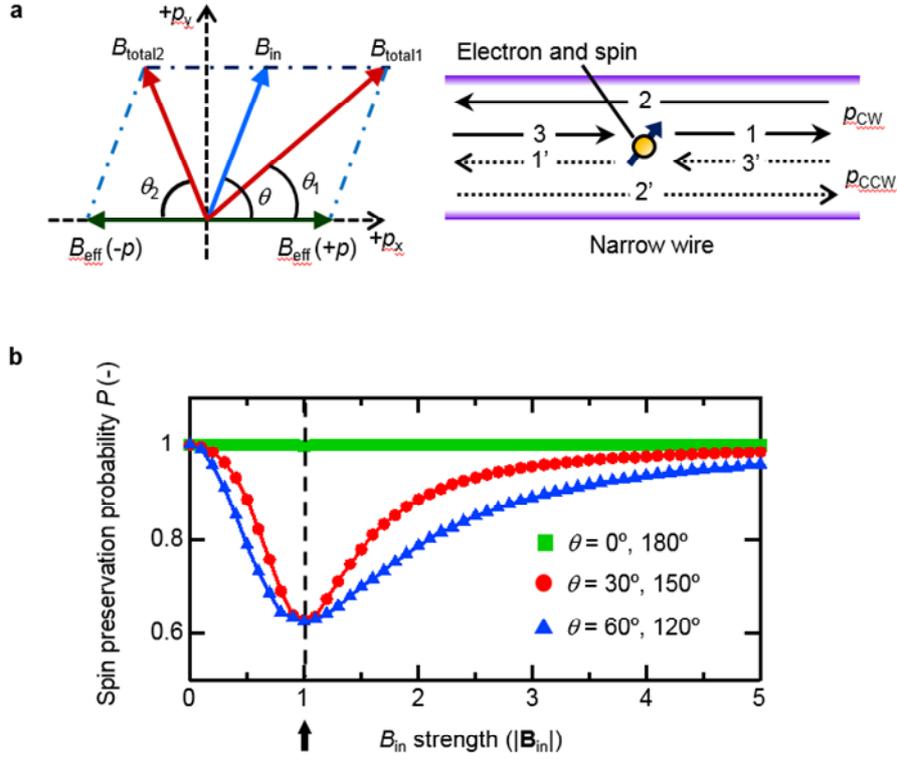

**Figure S2: Inference of spin relaxation from the toy model. a.** Schematic images of the modeled processes. In the quasi-one dimensional quantum wire the effective magnetic field $B_{\text{eff}}$ is uniaxial regardless of the electron momentum. $B_{\text{total1}}$ and $B_{\text{total2}}$ are defined by a superposition of $B_{\text{eff}}$ and the magnetic in-plane field $B_{\text{in}}$, resulting in total fields of $\mathbf{B}_{\text{total1}} = \mathbf{B}_{\text{eff}}(+p) + \mathbf{B}_{\text{in}}$ and $\mathbf{B}_{\text{total2}} = \mathbf{B}_{\text{eff}}(-p) + \mathbf{B}_{\text{in}}$ respectively. In the reference frame of the moving electron spin precesses in the two paths associated with clockwise (CW) and counterclockwise (CCW) spin rotation respectively. Both paths interfere at the point of origin. **b.** Dependence of the resulting spin preservation probability $P$ due to the interference in different direction $\theta$ (the angle between $B_{\text{in}}$ and $B_{\text{eff}}$). We fix $|\mathbf{B}_{\text{eff}}| = 1$ here. The shown values of $\theta$ include $\theta = 0º$, 180º (green rectangular), $\theta = 30º$, 150º (red circle) and $\theta = 60º$, 120º (blue triangle). Lower probability of spin preservation indicates lager spin relaxation, induced by precession around the axis defined by $B_{\text{total}}$. An increase of spin relaxation gives rise to a diminished WL signal.

Since the WL amplitude in in-plane field is suppressed together with the spin relaxation processes, we anticipate that the WL amplitude should be most suppressed when $|\mathbf{B}_{\text{in}}| \approx |\mathbf{B}_{\text{eff}}|$. In order to confirm this point we further conduct numerical simulations using the recursive

Green's function scheme introduced in paragraph 1. The $|\mathbf{B}_{in}|/|\mathbf{B}_{eff}|$ dependence of the WL amplitude is investigated in [100] quasi-one dimensional wires. In this simulation, we choose the following numerical parameters: Phase coherent length $L_\varphi$ = 1000 nm, spin precession length $L_{so} = \hbar^2/2m^*\alpha$ = 166 nm, mean free path $L_{el}$ = 17 nm and wire width $W$ = 100 nm. As the result shown in Fig. 6a indicates, minimum amplitudes of WL appear when the $|\mathbf{B}_{in}|/|\mathbf{B}_{eff}|$ ratio is around 1.0 for different $\theta$ angles ($\theta$ = 60º, 120º, 150º), The deviation of the curves recorded for $\theta$ = 30° is in line with the fact that large number of propagating orbital channels is not supported in the numerical scheme. As a consequence of this, non-universal effects on the individual mode cannot be negligible in calculation, which explain the observed deviation. Hence, it is possible to estimate the approximate $B_{eff}$ strength by using the relation of $|\mathbf{B}_{in}|/|\mathbf{B}_{eff}| \approx 1$, for which we observe the minimum WL amplitude.

Moreover, when we convert the deduced $|\mathbf{B}_{eff}|$ (the unit is tesla) in [100] wire into the absolute values of $\alpha$ and $\beta$ (the unit is $eV \cdot m$), we use the following equation,

$$\left|\mathbf{B}_{eff}\right| = 2\sqrt{\alpha^2 + \beta^2} k_F \big/ g\mu_B, \quad (S16)$$

Here, $g$ is the effective Landé g-factor and we assume $g$ = 3.5 (Ref. 40). $\mu_B$ is the Bohr magneton and $k_F$ is the Fermi wave number in 2DEG. By applying equation (S16) together with the estimated $|\mathbf{B}_{eff}|$ and the $\alpha/\beta$ ratio obtained from WL anisotropy measurements in the [100] wire, $\alpha$ and $\beta$ can be deduced.